# Model-Driven Engineering of Manufacturing Automation Software Projects - a SysML-based Approach

Birgit VOGEL-HEUSER, Daniel SCHÜTZ, Timo FRANK, Christoph LEGAT

Institute of Automation and Information Systems, Technische Universität München

**Abstract**: This paper comprises a SysML-based approach to support the Model-Driven Engineering (MDE) of Manufacturing Automation Software Projects (MASP). The Systems Modeling Language (SysML) is adapted to define the SysML-AT (SysML for Automation), a specialized language profile that covers (non-)functional requirements, corresponding software applications and properties of proprietary hardware components. Furthermore, SysML-AT supports an automated software generation for run-time environments conforming to IEC 61131-3. A prototypical tool support was realized for adapted SysML Parametric Diagrams (PD) inside an industrial automation software development tool. Coupling the model editor and online data from the provided run-time environment enables direct debugging inside the model. The approach was evaluated by several case studies and additional usability experiments. With the latter, the suitability of the MDE approach for future users was proven.

## 1. Introduction

Modern trends in manufacturing are defined by mass customization, high variability of product types, and a changing product portfolio during the life cycle of a manufacturing system [1, 2]. These trends imply more complex manufacturing systems [3], which support changes in the physical layout and technical updates.

Along with the complexity of the manufacturing system, which represents a special class of mechatronic system [2, 4], the complexity of the Manufacturing Automation Software Projects (MASP), which include information regarding applied automation hardware, is rising. Since the proportion of system functionality that is realized by software is increasing [5], concepts for supporting automation engineers in handling this complexity are strongly required.

The use of IEC 61131-3 run-time environments is currently the state of industrial practice in most manufacturing systems and will be the standard in the next 5-10 years according to [6]. As mentioned by [7], Programmable Logic Controller (PLC) platforms, which include soft-PLCs implemented on industrial PCs, are predominantly used in industry although other PC-based solutions exist. Consequently, this paper focuses on automation systems controlled by PLCs or soft-PLCs (both referred to as PLC henceforth), which use an IEC 61131-3 compliant run-time environment. Thus, the approach considers the PLC computing paradigm, i.e. the cyclical execution of the implemented code, that is reflected in the IEC 61131-3 standard. By focusing on PLCs, in contrast to embedded systems in general, vendor-specific platforms with limited configuration capabilities and fixed hardware characteristics are used. For this reason, automation hardware and its properties can be abstracted more strongly than it would be the case for general embedded systems/controllers.

Modeling approaches for software applications [8, 9, 10, 11] are a key issue for handling software complexity. Consequently, model-driven approaches have been developed, e.g. [9, 10, 12, 13], which address different systems (e.g. product automation, manufacturing automation), modeling languages (e.g. adapted or newly developed languages) and design phases (e.g. systems engineering with later code generation). For model-driven engineering (MDE) of manufacturing systems in general, three initial solutions may be chosen: (a) the adaptation of a widely accepted modeling language, e.g. Unified Modeling Language (UML) [14], with a standardized meta-model, diagrams and notations; (b) the development of a self-tailored modeling language; and (c), the combination of different modeling languages for different phases or disciplines. However, the coupling of different modeling languages for different phases and disciplines implies (multiple) bidirectional model transformations. Self-tailored modeling languages are faced with lower acceptance and increased adaptation costs by practitioners while being strongly application-specific.

With an existing modeling language (e.g. UML) that provides extension mechanisms, specialized language dialects can be defined by creating profiles. To increase the acceptance regarding a novel approach, easily applying and reproducing its concepts have to be enabled for other researchers by providing supplementary material [15]. The adaptation of a widely used modeling language for a MDE approach can additionally increase a concept's reproducibility since other researchers may already be provided with basic tool support for the modeling language. More importantly, integrations with other approaches, e.g. model-based test case execution [16], that extend the same base language, are simplified. For these reasons, the first alternative, the adaptation of a widely accepted modeling language, was chosen. Since differences between the implementation of the IEC 61131-3 standard by different PLC vendors exist, the MDE approach has to abstract these different implementations by an independent model and provide transformations to vendor specific implementations. Consequently, by applying the Model-Driven Architecture (MDA) paradigm [17] of the Object Management Group (OMG), a platform-independent model (PIM) to generalize different platform-specific models (PSM) of MASPs and a transformation between both has been developed.

This paper comprises a MDE approach for the domain of PLC-based manufacturing automation software projects (MASP). The novelty of the concept presented in this paper is based on the realization and integration of the following four key aspects into a consistent MDE approach for this domain:

- explicit modeling of functional and non-functional requirements [18, 19, 20, 21, 22] of PLC-based MASP
- modeling automation hardware architecture including proprietary components' characteristics [23, 24, 25]
- support mapping from abstract machine functions to real hardware-based functionality [18, 26, 27, 28]
- implementation supported by code generation [8, 9, 10, 11] and monitoring of executed software in the model

The SysML-based code generation support and the support for debugging monitoring of executed software inside the model was seamlessly integrated into a market-leading development tool (CODESYS V3) for IEC 61131-3 applications by extending prior works of the authors that developed a tool support for UML modeling inside this development environment. As mentioned in [29], "*Model-based debuggers are widely missing*" in software development. This also holds for the domain of MASP, until now. The approach was evaluated by case studies and also with intended users showing its strength compared to approaches that use existing programming concepts.

The remainder of the paper is structured as follows: In the next section, detailed requirements regarding support for the development of MASPs are derived. Subsequently, related work is discussed regarding the derived requirements. In section 4, an application example presents the SysML-based modeling approach, the SysML-AT, in detail, ranging from requirements modeling to software generation. The software generation is presented in section 5. The evaluation of the approach is given in section 6, considering an experimental study that investigates the usability of the approach and the fulfillment of the imposed requirements. The paper concludes with a summary and an outlook on future works.

## 2. Requirements for MDE of MASPs

In the following, the aspects that have to be considered by the MDE of MASPs are derived in an order that represents the overall design flow of the proposed MDE approach, from requirements engineering to detailed software modeling (cf. Fig. 1). However, during the early requirement engineering phase, specifications are usually incomplete and even full of conflicts. Hence, the proposed MDE approach does not require a strict sequential order and supports also iterative design flows for model refinements. The following paragraphs present the requirements R1 to R5 of the MDE approach, in which the relevant aspects are defined.

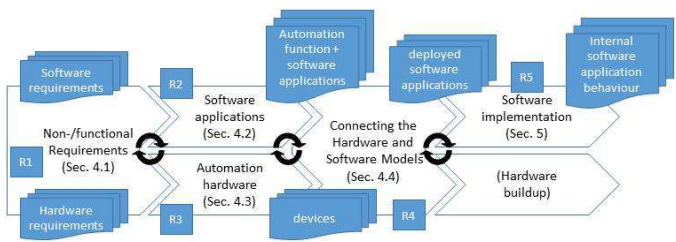

**Fig. 1: System design phases [30]**

### (R1) Modeling of (non-)functional requirements

The explicit modeling of requirements is an inevitable step in the development of embedded [19] and mechatronic systems [20]. The same holds for MSs, where functional as well as non-functional requirements, both resulting from the desired manufacturing process, have to be considered explicitly. While functional requirements are used to describe the demand for certain production tasks, non-functional requirements must be considered in order to describe real-time requirements of a closed-loop control or the required dependability of machine components [31], for example.

### (R2) Modeling machine functionality and software applications

A manufacturing system's functionality, i.e. the manufacturing process, is realized by a variety of different functions (referred to as automation functions (AFs) henceforth) like the separation of material, its transportation or forming. Hence, a successive identification of AFs during the decomposition and refinement of (functional) requirements has to be supported. These AFs are in turn realized typically by a (non-empty) set of implemented, programmatic building blocks, referred to as software applications (SA), which may be distributed on a variety of components. A detailed discussion on AFs and their distribution along production systems is given in [32]. Consequently, a modeling notation has to support the separation between (distributed) programmatic building blocks, i.e. SA, and a manufacturing system's or component's functionality, i.e. AFs, which are realized by compositions of SAs. Although the structure of SAs can be deduced during the refinement of requirement models, the developer needs to be supported by notation elements and model concepts (cf. R5) as presented in previous works [33] for manual, detailed descriptions of SA behavior.

### (R3) Modeling automation hardware and its characteristics

In order to realize the AFs, a manufacturing system usually contains a large number of sensors and actuators as well as a number of controllers (called nodes henceforth), which may be unknown in early design stages. Inside a MASP, nodes are characterized by their communication interfaces to each other, i.e. between different controllers or a controller and some sensors/actuators, which provide access to the physical plant for the implemented software [34]. The non-functional requirements, e.g. modularity aspects, are described in the requirement model of a MASP (see R1). A requirement regarding modularity would be necessary if, for example, a specific set of AFs needs to be realized by a single (mechatronic) component. To support the necessary deployment decisions of AFs to nodes, the corresponding non-functional properties of the hardware, e.g. available memory, have to be recognized as well.

**(R4) Connecting hardware and software models**

The deployed SAs need access to sensor values and control over the actuators that are connected to different nodes. Hence the connections between the SA's interfaces and a node's hardware interfaces to sensors and actuators must be modeled. Since communication between SAs is necessary, the modeling of connections between their interfaces is needed, too. In case of applications deployed onto different nodes, communication connections between the SAs need to be assigned to hardware interfaces of particular nodes.

**(R5) Implementation, operation and maintenance**

To support the implementation of SAs and provide another benefit of the MDE approach, an automated generation of executable code as presented by [8, 9, 11, 35] has to be provided. Consequently, following OMG's MDA paradigm, the proposed MDE approach needs to provide transformation rules between the SysML-AT model as the PIM and the (vendor specific) IEC 61131-3 code as the PSM. As identified in [36], modeling of user defined control logic is required in addition to the application of predefined control blocks. Hence, this work focuses on the specification of user defined control logic.

By addressing the generation of IEC 61131-3 compliant code, a wide range of applications can be addressed. In later steps, the approach may be extended with additional PSMs to address other commonly used platforms, e.g. robot controllers. Due to the cyclical execution of PLC-based automation software, no program loops with an external event as termination condition, e.g. the activation of a sensor, are allowed for valid code. Furthermore, since the invocation order of SAs is crucial for implementing a defined behavior of a MS, a manual specification by the application developer has to be possible.

The consistency between the model and the implemented software has to be ensured, manual changes of the generated code must either be traced back to the model or be avoided by prohibiting manual changes of the generated IEC 61131-3 code. The latter option can be realized by integrating the support for software modeling and model transformation into one development tool for automation software. In order to avoid maintenance of both, the model and the program code, a direct coupling between PIM and deployed code (PSM), is required which enables debugging of the software inside the PIM.

## 3. Related Work

The rising complexity of mechatronic systems and the lack of appropriate methods and modeling approaches supporting their development have been recognized by many researchers.

In particular, the current practice of designing software after a detailed mechanical design is criticized in many works. Because of this practice, design flaws often have to be fixed by software [37] making its development more complex. Hehenberger et al. [38] recognize that the current sequential design practice hinders the finding of optimal designs for a mechatronic product. The solution proposed in [39] provides a functional layer from which detailed artifacts of various domain models can be derived. However, only the software architecture is generated (cf. R2); the modeling of user defined control logic and integrations with software development and run-time environments (cf. R5) are not provided.

There are modeling approaches for mechatronic components which are far more detailed. They are often based on formal languages, e.g. Bond Graphs, describing the physical aspects of components [40, 41, 42]. Other models, like QUASIMODO, describe the underlying network architecture in detail [43] or are specialized for a specific goal, like DECOS [44] for the development of dependable embedded software. DECOS implies a message based communication middleware [45]. SAs, called jobs, communicate via ports with read and write messages. This concept does not satisfy R5. Languages like AADL describe the interaction between the physical system architecture, the runtime architecture and the computer platform and "[...] *do not focus on the system as a whole*" [46]. Models necessary to develop MASP and the models for mechatronic component design differ in size and level of abstraction. In MASP proprietary hardware descriptions are included, however, complex physical models with high level of detail are of less interest.

Numerous works address the development of specialized modeling approaches that support the development of various aspects of manufacturing systems. Vepsäläinen et al. [36] present the AUTOKON approach that covers several aspects of the approach presented here (cf. R2 – R4). However, since (non-)functional requirements are developed but only imported and elaborated, requirement R1 is not fulfilled. The results of Vepsäläinen et al. show that industrial practitioners are missing possibilities to specify user defined or application specific control logic [36] as well as possibilities to explicitly specify the execution order of Function Blocks [36]. Hence, requirement R5 is not fulfilled as well. The integration of various domain-specific views using SysML to support systems engineering has been proposed, e.g. in [47, 48]. These works develop language profiles of SysML that support the modeling of (non-)functional) requirements, e.g. real-time requirements, and test cases to verify the requirements [48] (cf. R1) as well as the detailed behavior modeling [49] (cf. R2 – R4). However, since these works require transformations to external simulation environments (MATLAB/Simulink [49] or Modelica [47]) to execute the models, a direct coupling of the SysML models with the executed IEC 61131-3 code (as required by R5) is not provided. Approaches like [50, 51] generate IEC 61131-3 code based on Matlab models but do not cover the requirements R1 and R4. In the MEDEIA project [52], a framework that applies the MDA approach for industrial automation control systems was developed considering both, event-based and cyclic PLC platforms. Therefore, code overhead is required when transforming e.g. modeled event-based behavior to IEC 61131 [53]. The industrial evaluations of the MEDEIA approach presented in [53] indicate that the approach lacks appropriate application debugging for industrial practitioners and, hence, does not fulfill requirement R5.

Thramboulidis' approaches [5, 54] are also based on SysML and support the modeling of functional and non-functional requirements for the development of domain-specific components, i.e. the software for mechatronic systems (cf. R1 – R4). However, these works and further elaborations [10] solely model the software structure, refer to standard tools for detailed implementation, and lack coupling between the models of the software and run-time environments (cf. R5). Estevez et al. [13] provide a coupling of models from various domains involved in manufacturing systems engineering based on XML schemes and the approach from [11]. Although the structure of the software can be described by modeling Function Blocks [55] and their software interfaces (cf. R2 and R4), the implementation of SAs require model exports to PLCOpen [56], a standardized interchange format for PLC programs [57]. AutomationML [58], a generic data exchange format based on XML to support MDE for MS, utilizes PLCOpen for representing control software, too. Due to the different vendor-specific interpretations of the IEC 61131-3 standard, a direct im-/export of AutomationML or PLCOpen respectively requires additional effort [53]. A direct integration of PLC software development and run-time environments, directly linking the model with the executable code (cf. R5) is not provided. Adaptations of UML and SysML to describe different aspects of mechatronic systems at different hierarchical levels provide the basis for the works presented in [9, 59, 60]. Secchi et al. [59] and Bonfé et al. [41] demonstrate the advantages of supplying object-oriented models with a formal basis in order to apply methods for verifying modeled system requirements [61] (cf. R1). The modeling approaches used in these and other works [9, 60] provide the integration of software models and physical models (cf. R2 – R4) of mechatronic components into one single component description using a single consistent syntax. To generate the detailed code for modeled SAs, design patterns and transformation rules are proposed in [9]. Although these works allow state chart models to be transformed into executable PLC code, a direct integration of the software model and executed code is not provided, and requirement R5 is violated by those approaches.

Aside from focusing on implementations in IEC 61131-3, research is being conducted that addresses event-driven implementations conforming to the IEC 61499 standard [62, 16, 42, 43, 63, 28]. The system engineering framework proposed by Hirsch et al. [63, 28] is based on SysML, IEC 61499 code and abstraction levels for the models that are similar to the ones in the approach presented here. Although the approach from [28] does not extend the SysML meta-model, the definition of specialized language profiles is considered to be useful for larger software projects (cf. [28]). The system engineering framework further allows the modeling of requirements (cf. R1) and tracing of requirements to artifacts for the system elements (cf. R2). However, the modeling of hardware access is not included in the approach of Hirsch et al. [63, 28] and debugging directly inside the SysML model is not provided. Hence, the requirements R3 – R5 are not fulfilled. Nevertheless, in addition to the approach presented here a formal verification of the IEC 61499 code, which can be generated from SysML models inside the framework is provided.

Dubinin et al. [64] generate software according to IEC 61499 from UML-based models. Model transformations between IEC 61499 and MATLAB/Simulink models have been presented by Vyatkin et al. [62] and Yang et al. [8] in order to support both, the transformation of software to MATLAB/Simulink for verification and the transformation of controllers designed using MATLAB/Simulink into executable software. However, the model development uses a separate modeling environment that is not integrated into the software development (cf. R5) and hence is not coupled with the run-time environment of the generated code. Furthermore, although solutions for integrating IEC 61499 [65] run-times on state of the art industrial controllers exist [66], according to [7]: "*IEC 61499 has a long way in order to be seriously considered by the industry*".

To describe automation hardware devices (cf. R3) several languages like EDDL [67], FDT [67] or FDCML [68] for describing field devices and networked architecture systems in a detailed way exist. Such Languages are often used by vendors to provide standardized documentations and installation instructions for their devices. However, they describe a device's parameters, interfaces and communication aspects in detail and, hence, require information that does not exist in the early design phases. For the MDE approach proposed in this paper only a limited set of relevant characteristics of proprietary hardware for MASP development will be focused (cf. R3). However, the approach may be extended by these device descriptions to address detailed hardware modeling and architectural development.

Summarizing this section, numerous promising approaches exist for the MDE of mechatronic systems and manufacturing systems in particular. The use and adaptation of object-oriented modeling languages like UML and SysML is widespread [10]. However, although existing concepts partly fulfill the imposed requirements, an approach that covers all necessary aspects does not exist yet.

## 4. Automation System Development (R1-R4)

This section presents the phases corresponding to R1 – R4 of the MDE approach for MASP. As a basis, the Systems Modeling Language (SysML) [69], which is a further development of UML [14], was adapted to form the SysML for Automation (SysML-AT). SysML has been developed to support systems engineering and to provide model artifacts, e.g. for physical interfaces (*Flowports*), not covered by UML [5].

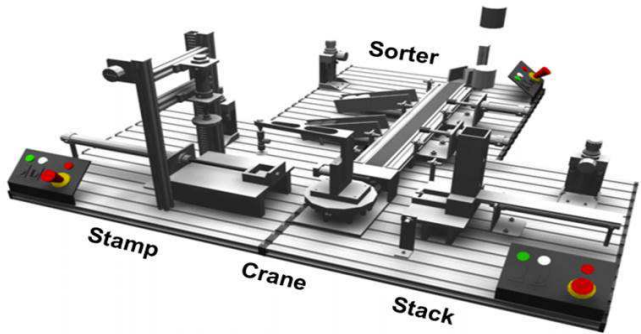

**Fig. 2: Application example Pick&Place Unit (PPU) [70]**

Most important, SysML covers the modeling of functional requirements and corresponding software applications as well as the modeling of non-functional requirements [71], e.g. modularity, and properties of integrated hardware components. Functional requirements describe the functional behavior desired at a later point; non-functional requirements constrain architectural decisions.

The laboratory manufacturing system called PPU [70], which will be used as a running application example, is an integration of four components (see Fig. 2), i.e. a stamp, a crane, a stack and a sorter. The stack stores and separates cylindrical chumps by extending a pneumatic cylinder. The crane is installed on top of a turntable and includes a pneumatic cylinder with a gripper that enables picking and placing chumps at different positions. Inside the stamp, chumps can be stamped by a pneumatic cylinder. The overall laboratory manufacturing system enables that chumps can be picked from the stack by the crane and transported to the stamp. After the chumps have been stamped, they can be forwarded to the sorter. To control the PPU according to the specified requirements, several sensors and actuators are integrated, e.g. a motor that rotates the base of the crane. Although there is no need for two PLCs from the size of the algorithms to be implemented, two PLCs with IEC 61131-3 compliant run-times are provided to demonstrate the deployment of the software according to R3 and R4.

In the following subsections, the MDE approach based on SysML-AT is presented ordered by the requirements R1 – R4. Fulfillment of requirement R5 is presented in section 5.

### 4.1 Modeling of (non-)functional requirements (R1)

The objective of a manufacturing system, i.e. the behavior that is required to realize a manufacturing process, can be described as a set of requirements, consisting of functional and non-functional requirements (refined during the engineering process). Since requirement specifications in the early requirement engineering phase are usually incomplete, the use of SysML enables their refinement throughout (later) design phases.

To specify functional requirements, the basic concept *Requirement* of SysML, providing a unique name and a text-based description, is used. By using the *Refine* relation, requirements can be decomposed and the description of a hierarchical requirement structure is provided.

The *Validity* relation can be used to connect functional requirements to corresponding modeled components of a manufacturing system. In contrast to functional requirements, the number of existing non-functional requirements is limited [72]. For this reason, according to other existing approaches, non-functional requirements can be described by various properties (represented as type-value pairs). Hadlich et al. [73] showed the possibility of assigning properties to requirements and other manufacturing system components. Non-functional requirements can also be connected to other notation elements using the *Validity* relation. Information on already modeled mechatronic components that fulfill a subset of the identified requirements can be integrated into a MASP model.

*Application example*

The crane of the PPU has to transport chumps from the stack to the stamp. This functional requirement can be subdivided into *angle detection* and *motor activation* (cf. Fig. 3). For the functional requirement *motor activation* an electrical drive (*M*) for the crane may be integrated into the design as an already designed mechatronic component.

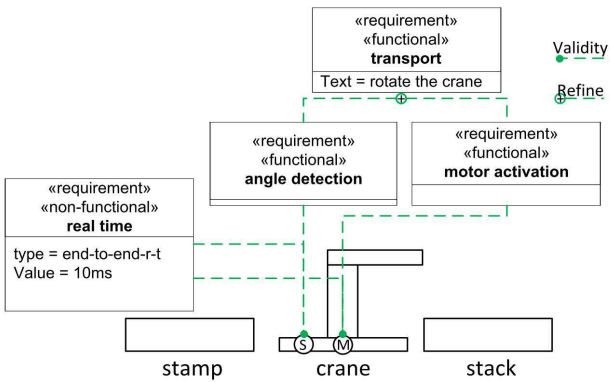

**Figure 3: Exemplary notation for 'requirements'**

Inside the model, this information is given by the non-functional requirement *real-time* and defined value for its property *end-to-end response time*, which reflects the maximum time between a change of the sensor signal (*S*) and the actuator (*M*) being activated [74]. The validation of a manufacturing system's design in early stages regarding this type of non-functional requirement [73] is described in [75].

## 4.2 Modeling machine functionality and software applications (R2)

The decomposition and refinement of functional requirements as described in the previous section result in the definition of automation functions (cf. R2). A correspondingly named concept and notation element *AF* is introduced in SysML-AT in addition to the concepts of SysML that encapsulate the complexity within a MASP model, e.g. by hiding the internal structure. AFs can be linked by several connections to define i.e. required data exchange. The likewise introduced concept and notation element for software applications *SA* details an AF and includes the functional behavior. Interfaces of a SA are represented by the SysML concept *Port*. In subsequent steps, the internal behavior of SAs can be specified by Parametric Diagrams of SysML-AT (cf. section 5).

*Application example*

To realize the requirement *angle detection*, two AFs, *angle measurement* and *safety shutdown*, can be added (cf. Fig. 4).

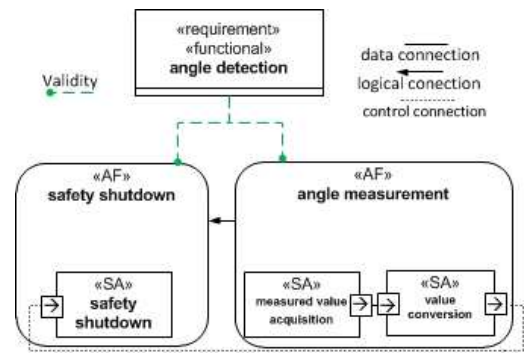

**Figure 4: Exemplary notation for 'software'**

The *safety shutdown* AF has a logical connection to the *angle detection* AF and realizes the monitoring of angle constraints. The *angle measurement* AF is further detailed by the SAs *measured value acquisition* and *value conversion*. The connection between these two SAs is a cyclically exchanged measurement value. The logical connection between the two AFs is later detailed by the control connection between the SAs *value conversion* and *safety shutdown*.

## 4.3 Modeling automation hardware and its characteristics (R3)

In this phase of the MDE approach, the automation hardware (basic characteristics) is derived from the imposed requirements and the SAs that have to be implemented. PLC-based automation software is executed mainly in vendor-specific run-time environments. Therefore, only a subset of relevant aspects is required. Corresponding notation elements (*sensor*, *actuator*, and *node*) based on the SysML stereotype *Block* are provided by SysML-AT. When a concrete hardware is chosen, device descriptions like EDDL can be used to complete its properties. Comparable to SAs, input and output ports can be added to these concepts for hardware elements to describe their interfaces.

To provide additional information about required hardware, depending on the device type, predefined properties can be added [73]. Properties relevant for hardware configuration like *type* (e.g. angle sensor), *bus type* (e.g. PROFINET [76, 77]) and *bus address* can be considered. To suggest which properties are robust with regard to the execution time, the SA's run-time environment must be defined in advance. Therefore properties like the PLC cycle time and the available memory space can also be added.

*Application example*

In the PPU, one motor and one angle sensor are required at the crane. According to imposed requirements, the execution time for the *measured value acquisition* AF must be smaller than 10ms.

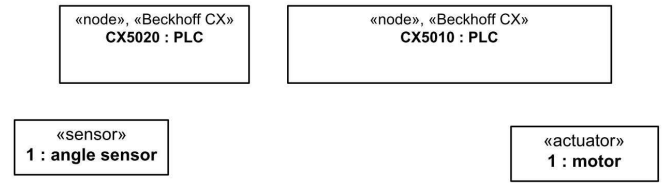

Fig. 5: Exemplary notation for 'hardware'

Consequently, as shown in Fig. 5, two nodes are used to separate one fast PLC (Beckhoff CX5020) and one slow PLC (Beckhoff CX5010).

## 4.5 Connecting hardware and software models (R4)

Once the SAs and nodes (including sensors and actuators) have been defined, these elements have to be mapped to each other. Depending on the non-functional requirements, e.g. real-time as given in the example above (execution time), an appropriate run-time environment for each SA has to be chosen. By allocating (deploying) the modeled SAs to nodes, it is specified on which type of run-time environment the particular SAs have to be deployed.

With the interaction between the SAs and their deployment onto nodes being defined, necessary communication between the different nodes can be derived. By the modeling of *Ports*, the information on the specific communication of a particular node is separated from the SA that may be deployed to it. Hence, the SAs are abstracted from node specific communication and can be reused inside the model and deployed to different nodes.

Furthermore, by specifying a vendor for the nodes and a bus type, e.g. a (vendor-specific) real-time Ethernet [76] fieldbus, it is defined which proprietary communication and synchronization mechanisms are used for node to node communication. In the proposed MDE approach, communication mechanisms are not described in detail but referenced and only the information necessary to implement the desired proprietary communication mechanism is considered. Subsequently, the connections between the software interfaces of SAs (input/output variables) and the hardware interfaces of nodes can be derived (cf. R4).

By this step, the complete software architecture is defined. This includes the SAs, their interfaces and interaction as well as sensors and actuators. As complex systems include a lot of connections the prototypical tool is extended by a context sensitive connection highlighting and filtering, i.e. by clicking on one node only the connected elements are displayed.

*Application example*

For the PPU it is assumed that the *measured value acquisition* SA requires an execution time of 3 ms and the *value conversion* SA is not constrained and contains complex mathematical calculations. The two SAs are deployed on two different PLCs from the PLC vendor Beckhoff (cf. Fig. 6). Hence, the modeled logical connection has to be realized by communication between the two nodes. In the PPU (cf. Fig. 6), the proprietary publisher/subscriber mechanism for transferring variables between Beckhoff PLCs is used to realize this SA to SA communication. To implement the communication using this mechanism, information on the exchanged variables, i.e. their name, data type and accessibility (subscriber/read or publisher/write) as well as on the nodes exchanging these variables is necessary.

Consequently, the modeling approach introduces the stereotype "Beckhoff CX", which enables describing the address of a Beckhoff specific node by its AMSNetId (cf. Fig. 6). The name and data type of the transferred variables are considered by the corresponding properties of the modeled ports (e.g. 'Input' and 'Output' in Fig. 6).

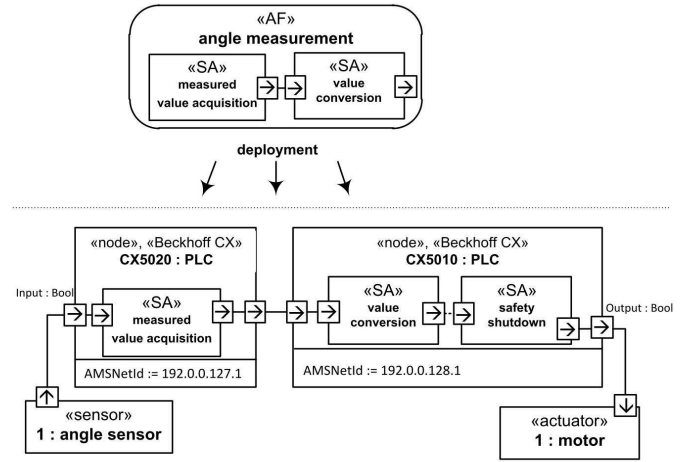

**Fig. 6: Example and notation for 'deployment'**

With the direction of a port the accessibility of a variable is defined and a corresponding publisher or subscriber variable can be created, e.g. from an outgoing port a publisher variable of a node can be created. In a last step, the information on which publisher and subscriber variables (modeled as *Ports*) are linked can be extracted from the *data connections* which are based on the SysML concept *BindingConnector*. For the definition of compatibility between ports, the syntax and semantics defined inside the SysML specification are applied: "*The two ends of a binding connector must have either the same type or types that are compatible*" [78].

## 5. Implementation, operation and maintenance

This section presents the phase corresponding to R5 of the MDE approach for MASP. To fulfill requirement R5, the language profile SysML-AT enables the modeling of SAs and generation of IEC 61131-3 code based on the SysML Parametric Diagram (PD). For modeling and generating SAs with discrete behavior, a tool support for adapted UML State Charts has been developed in previous research projects of the authors [35] and is now commercially available inside Version 3.5 of the CODESYS programing environment.

These previous works are complemented with a support for modeling SAs with continuous behavior based on the PD proposed in this paper.

In order to integrate the PD-based software models with the run-time environments, like in the previous works, a model editor was integrated into the IEC 61131-3 development environment. The expressiveness of the realized PD editor regarding program loops, e.g. Do-While, is equal to the languages Ladder Diagram (LD) or Function Block Diagram (FBD) of IEC 61131-3 and Continuous Function Chart (CFC). These languages do not provide the use of program loops and, thus, per se ensure that the described code can be executed cyclically. The direct debugging of the model was enabled by coupling the model editors with online data from the run-time environment.

The following subsections will present relevant language concepts of IEC 61131-3. Subsequently, the language concepts of the SysML-AT are presented and the code generation for PDs is described. The integration of corresponding editors is presented in section 5.4.

## 5.1 Language concepts of the IEC 61131-3

IEC 61131-3 provides several language constructs for implementing automation software: Function Blocks (FB), Variables (VAR) and Functions (FC) [79].

A FB is the main construct for encapsulating code. It is divided into a declaration part and an implementation part. Within the declaration part, other FBs can be instantiated. Additionally, a FB can contain declarations of variables with primitive (e.g. INT, BOOL) or user defined data types. These variables can be declared as local variables (VAR) for storing calculation results or as input/output variables (VAR_IN, VAR_OUT) for providing a signal interface of the FB.

FCs consist of a declaration part and an implementation part but cannot be instantiated and, thus, cannot contain persistent variables. The output signal interface of a FC is constrained to one return value of the calculation, which may be a primitive or user defined data type.

The implementation part of FBs and FCs can be implemented using the languages standardized in the IEC 61131-3. The languages provide constructs for defining sequences of value assignments between variables (VAR), constant parameters and signal interfaces (VAR_IN, VAR_OUT) of FBs and FCs as well as the invocation of FB instances and FCs. Subsuming, to enable the software generation, a SysML model has to contain equivalents for the following language constructs:

- Function Block Instances (FB Inst.)
- Local Variables (VAR)
- Input Variables (VAR_IN)
- Output Variables (VAR_OUT)
- Function Block Invocations (FB Inv.)
- Function Invocations (FC Inv.)
- Function Blocks (FB)
- Functions (FC)
- Value assignments (VA)

In the following section, the parts of the language profile SysML-AT for generating IEC 611313-3 code are derived from stated requirements for model-based code generation.

## 5.2 Detailed software models in SysML-AT

For defining the SysML-AT, the original meta-model of the Unified Modeling Language (UML) [80] and the adaptations to it introduced by SysML [78] are further extended by several stereotypes. The provided adaptations can be divided into four major aspects: Instances, Classifiers, Flows and Restrictions. The 'Instances' aspect summarizes model elements that describe objects, e.g.

instances of FBs. The aspect 'Flows' summarizes model elements that describe information connections, i.e. value assignments between e.g. variables. The 'Classifiers' aspect outlines model elements that describe classes of elements, i.e. FBs or FCs. The constraints of the stereotypes are summarized under 'Restrictions'.

*Instances*

The basic elements of UML used for the proposed language profile are the meta-classes *Property* (describing instances) and *Port* (describing signal interfaces) (cf. Fig.7).

SysML extends these meta-classes to distinguish between *PartProperties*, *ValueProperties* and *ConstraintProperties*. PartProperties represent complex decomposable instances, whereas ValueProperties represent simple parameters with a primitive, e.g. Integer, or user defined data type.

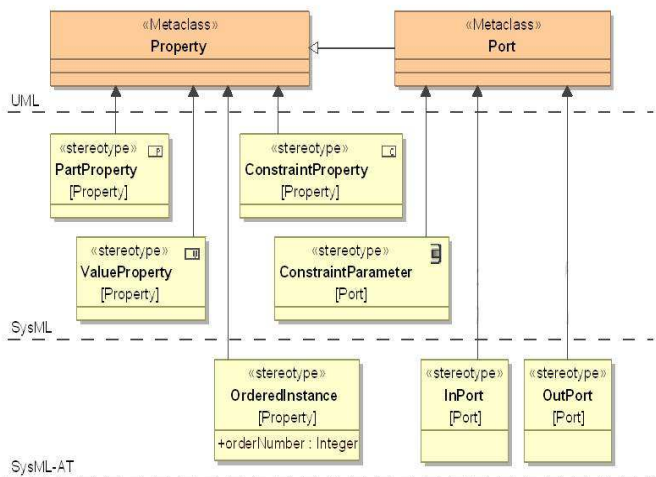

**Fig. 7: Profile diagram of SysML-AT -- Instances**

With the stereotype *ConstraintProperty* objects referring to a simple algebraic rule are modeled. The presented approach requires that these rules are expressed as (multiple) statements of the Structured Text (ST) programming language standardized in IEC 61131-3. The stereotypes *InPort* and *OutPort* introduced in SysML-AT are derived from the meta-class *Port* and represent special ValueProperties that are used to describe the signal interface of a Property. The mapping between these SysML-AT elements and IEC 61131-3 is given in Table 1.

**Table 1: Mapping SysML-AT to software – Instances and Invocations**

| SysML-AT | IEC 61131-3 |
|---|---|
| PartProperty | FB Instance |
| ValueProperty | Variable |
| Port<br>- InPort<br>- ConstraintParameter + InPort | Input Variable:<br>- Function Block (FB)<br>- Function (FC) |
| Port<br>- OutPort<br>- ConstraintParameter + OutPort | Output Variable:<br>- Function Block (FB)<br>- Function (FC) |

The instances of FBs can be modeled as PartProperties, and local variables as ValueProperties. The signal interface of a FB is modeled using the stereotypes InPort and OutPort; for the signal interface of a FC the stereotype ConstraintParameter is applied additionally.

**Table 2: Mapping from SysML-AT to software – Invocations**

| SysML-AT | IEC 61131 |
|---|---|
| ConstraintProperty + OrderedInstance | FC Invocation |
| PartProperty + OrderedInstance | FB Invocation |

The stereotype *OrderedInstance* is introduced for invocations of FCs and FB Instances. Since the invocation order is crucial for the behavior of the overall software, this stereotype specifies the tag *orderNumber* to define an invocation order. By applying the stereotype OrderedInstance, a defined sequence of invocations in the software can be modeled (see Table 2).

*Classifiers*

In a SysML model, *Blocks* are used to describe classifiers of complex, decomposable objects (PartProperty) and *ConstraintBlocks* are used to describe classifiers of simple objects (ConstraintProperty) that are constrained by an algebraic rule (*Constraint*).

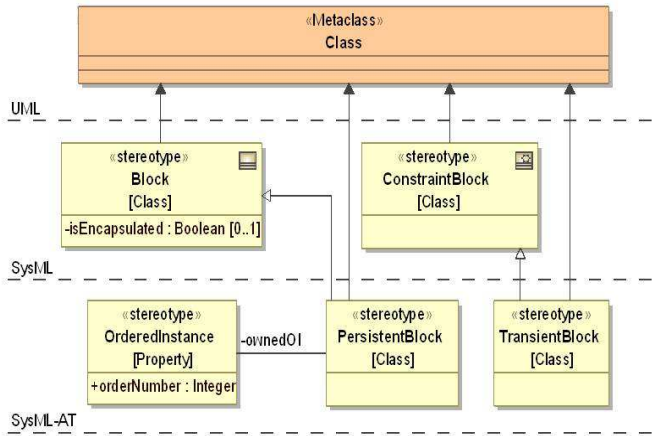

**Figure 8: Profile diagram of SysML-AT – Classifiers**

To model FBs and FCs, the stereotypes *PersistentBlock* and *TransientBlock* are derived (cf. Table 3). Inherited from the meta-classes of UML (cf. Fig. 8), PersistentBlocks and TransientBlocks are able to contain the meta-elements that have been defined to represent the elements of FB Instances and invocations as well as FC invocations in the previous subsection.

**Table 3: Mapping SysML-AT to software – Classifiers**

| SysML-AT | IEC 61131 |
|---|---|
| PersistentBlock | Function Block |
| TransientBlock | Function |

To fully comply with these software elements, the stereotypes have to comply with the restrictions that are presented in the corresponding subsection.

*Flows*

To interconnect objects like ValueProperties or Ports, a stereotype is derived from the meta-class *InformationFlow* (cf. Fig. 9). Elements of that meta-class already contain the feature to specify the source element and target element of the modeled connection.

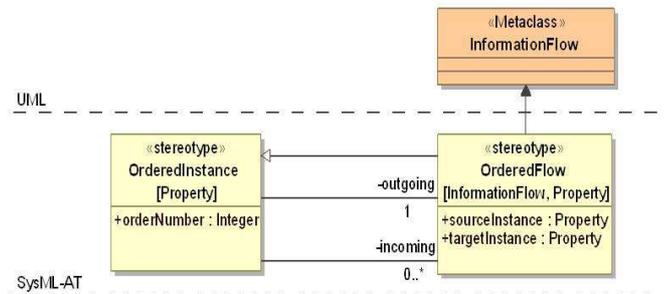

**Figure 9: Profile diagram of SysML-AT – Flows**

However, since source and target of an InformationFlow refer to a classifier, instance specific connections cannot be expressed. Hence, the stereotype OrderedFlow contains the tags *sourceInstance* and *targetInstance*, which allow the specification of an instance for which the modeled connection is valid.

**Table 4: Mapping SysML-AT to software – Flows**

| IEC 61131 | SysML-AT |
|---|---|
| Value Assignment | OrderedFlow |

Furthermore, since the assignment of variables for the software is modeled using connections of this stereotype (see Table 4) it contains the tag *orderNumber*, which indicates the place inside the code at which the assignment is to be generated.

*Restrictions*

To enable software generation from SysML-AT, the stereotypes need to be restricted by several constraints. The most important restrictions constrain the orderNumbers of the instances and flows which have to comply with a number of constraints: Firstly, in order to verify a distinct invocation sequence, the orderNumbers specified by the different instances inside a Block have to be greater than zero and different from each other. In contrast to instances of FBs (PartProperties), the inputs and outputs of FCs (ConstraintProperties) can only be manipulated (by OrderedFlows) during the invocation of a FC. Hence, the orderNumber of a ConstraintProperty also defines the positions of the connected OrderedFlows within the invocation sequence. Hence, the orderNumbers of Flows connected to a ConstraintProperty have to be equal to zero. For a FC, only one output variable, i.e. only the return value of the calculation, can be defined. Hence, the TransientBlocks of SysML-AT have to be constrained to just one OutPort as well.

**5.3 Software generation from Parametric Diagrams**

As provided by the presented approach, AFs are decomposed into SAs, which have to be implemented as FBs on PLCs to realize the modeled AF (see section 4.2). Each FB is described using a PersistentBlock and a Parametric Diagram (PD) that follows the language profile presented in the previous subsection. Fig. 10 exemplifies the modeling of the detailed behavior of a FB using a Parametric Diagram of SysML-AT.

To realize the generation of executable IEC 61131-3 code, a model-to-text transformation was realized using the MOF Model to Text Transformation Language (MOFM2T) [81], which is standardized by the Object Management Group (OMG). Using MOFM2T, transformation *modules* can be specified for meta-models that comply with OMG's Meta Object Facility (MOF) [82]. Inside such MOFM2T *module*, several textual *templates* can be specified in which the model elements' information is transformed into a textual representation by using the commands and syntax of MOFM2T in combination with the the Object Constraint Language (OCL) [83].

For transforming SysML-AT PDs to IEC 61131-3, a transformation module was defined that translates the meta-models of UML, SysML and SysML-AT. Inside this *module*, according to the defined mapping rules (cf. Tables 1 – 4), *templates* for the elements of a SysML-AT model have been defined that generate the textual representations of the declaration part of IEC 61131-3 FBs and FCs as well as their implementation part in IEC 61131-3 ST.

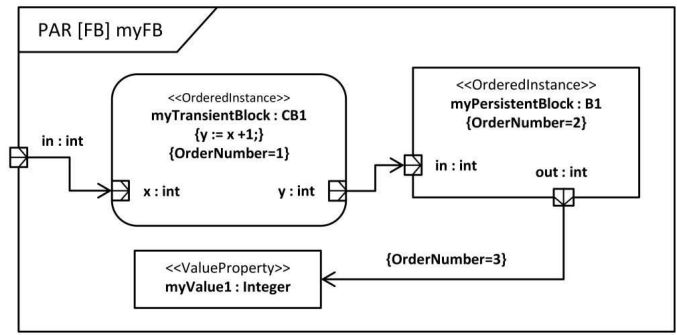

**Fig. 10: Example of a SysML-AT PD**

Consequently, for each PersistentBlock an IEC 61131-3 FB is declared and an IEC 61131-3 FC is declared for each TransientBlock. The input, output and local variables of FBs and FCs are generated according to the modeled InPorts, OutPorts and ValueProperties.

The implementation part of a FC is given by the algebraic rule of the corresponding TransientBlock as one or multiple statements in ST. Consequently, the template defined for the implementation part of IEC 61131-3 FCs extracts this textual information. The implementation part of a FB has to be generated from the information on the OrderedInstances of a PersistentBlock.

An excerpt of the template that generates IEC 61131-3 ST code from a set *ownedOI* of OrderedInstances is shown in Fig. 11. Inside the template, a MOFM2T *for-block* is used to iterate over the set *ownedOI* that is re-ordered according to the (*orderNumbers*) of the particular elements by integrating the OCL function 'sortedBy()' (see Fig. 11, a). Elements whose order number is equal to zero, as it would be the case for OrderedFlows that are connected to instances of TransientBlocks (cf. section 5.2, Restrictions), are filtered out by a MOFM2T *if-block*.

The template furthermore distinguishes between elements that are plain OrderedInstances (cf. Fig. 11, b) describing IEC 61131-3 FC and FB instance calls and elements that are OrderedFlows (cf. Fig. 11, d) describing value assignments between variables.

```
[comment: Process OrderedInstances by OrderNumber/]
[for (oi : OrderedInstance | pb.ownedOI->sortedBy(orderNumber))]
[if (oi.orderNumber > 0)]                                              a

[comment: Generate FC and FB-instance invocations /]
[let myoi : OrderedInstance = oi]                                      b
[if (myoi.type.eClass() = PersistentBlock)]
[myoi.name/]();
[/if]
[if (myoi.type.eClass() = TransientBlock)]
[TBOut(myoi)/] := [myoi.type.name/]([TBIn(myoi)/]);                    c
[/if]
[/let]

[comment: Generate Value Assignments/]                                 d
[let myoi : OrderedFlow = oi]
[myoi.targetInstance.name/].[myoi.informationTarget.name/]:=
[myoi.sourceInstance.name/].[myoi.informationSource.name/];
[/let]

[/if]
[/for]
```

**Fig. 11: Excerpt of the MOFM2T template for IEC 61131-3 ST code generation from SysML-AT models**

For generating textual representation of both, further information is extracted from the SysML-AT model using the MOFM2T syntax. The template invokes the templates *TBOut* and *TBIn* for processing the incoming and outgoing OrderedFlows of TransientBlocks, i.e. generation of value assignments between IEC 61131-3 variables and FCs (cf. Fig. 11, c).

```
1   myPersistentBlock.in :=
2   CB1(x:=in);
3   myPersistentBlock();
4   .myValue1 :=
5   myPersistentBlock.out;
```

**Fig. 12: IEC 61131-3 code resulting from applying the transformation shown in Fig. 11 to the PD depicted in Fig. 10**

Given the transformation depicted in Fig. 11, the PD presented in Fig. 10 relates to the code in Fig. 12.

## 5.4 Integrating SysML-AT within an IEC 61131-3 development tool

Based on the previous works, the presented approach for the modeling of SysML-AT Parametric Diagrams (PD) and software generation has been integrated into a widely used IEC 61131-3 development environment (CODESYS V3 [84]) which provides both, the editors for the standard languages as well as the run-time environments (soft-PLCs) to execute the software. In order to integrate the presented approach, an editor for SysML Parametric Diagrams has been realized (cf. Fig. 12) that can be used for the implementation of FBs in addition to the editors for the IEC 61131-3 languages.

Similar to the processing of code described in the IEC 61131-3 language Sequential Function Chart (SFC), the model transformation translates preprocessed SysML-AT PD into ST. To keep the consistency between SysML-AT models (the platform-independent model) and the generated software (the vendor platform-specific model), the latter currently can only be reviewed by the developer before it is sent directly to the compiler. Further elaborations of the approach will provide concepts for code design patterns similar to [85] to improve the readability of generated code as well as editing and maintaining of the generated code inside the IEC 61131-3 languages.

The implemented PD editor is able to use the same communication channels to the run-time environment as the native editors of CODESYS. Hence, an online view that provides online values for the executed automation software to the user was realized. Elaborations of the tool support currently integrate features to force variable values during the execution of the code and allow the specification of breakpoints and for debugging purposes.

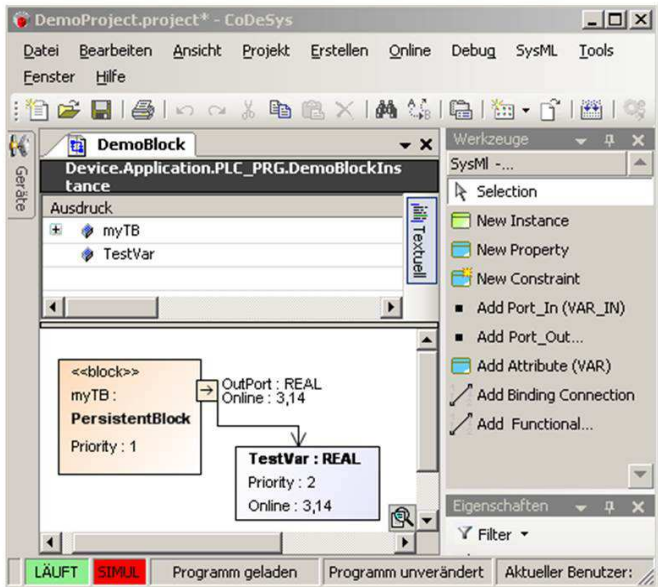

**Fig. 13: PD editor embedded in CODESYS (online view)**

With this integration of the SysML-AT modeling support, it is possible to develop SysML-AT models describing the software and to debug and maintain the deployed software.

## 6. Evaluation of the SysML-based approach

To evaluate the proposed MDE approach, its applicability for different manufacturing systems has been investigated as well as its usability for future application engineers and its applicability in industry. The following subsections present the conducted case studies followed by the usability evaluation with human subjects, i.e. students from mechanical engineering and the investigation of the industrial applicability by guided expert interviews.

### 6.1 Applicability of SysML-AT for different MSs

Several case studies proving the applicability of the proposed MDE approach have been conducted using different manufacturing systems, e.g. the application example from section 4 (see Fig. 2) and a larger scale manufacturing system (see Fig. 14). The latter incorporates plant sections for different production domains, such as fluid processing, logistics and handling robots.

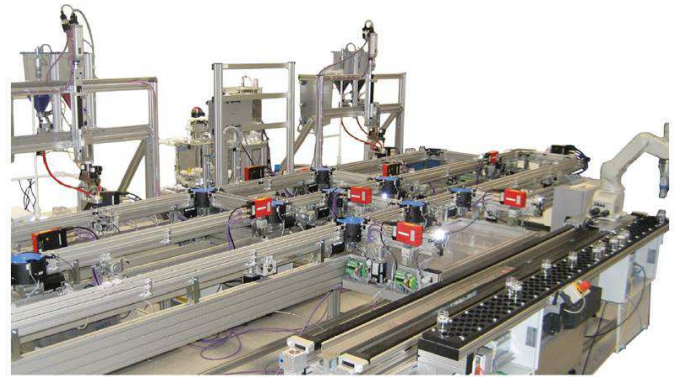

**Fig. 14: Hybrid Process Model -– Manufacturing system case study for evaluation**

The considered manufacturing systems contain centralized, decentralized and distributed hardware architectures built with PLCs from different vendors. The SysML-AT-based MDE approach including the generation of IEC 61131-3 software has been evaluated by four students' thesis projects, which proved the applicability of the MDE approach for MS. Furthermore, the developed tool support was applied as part of a one-week lab course, which was attended by 10 undergraduate students. Using the SysML-AT modeling approach and the integrated code generator, executable run-time code was generated and deployed onto the soft-PLC.

### 6.2 Usability of SysML-AT for future application engineers

To investigate the usability of the proposed MDE approach, usability experiments of the developed tool-support with human subjects have been conducted. The experiments included 36 students of mechanical engineering (3rd semester) at the Technische Universität München. Previous knowledge regarding IEC 61131-3 and SysML/UML was captured by several questionnaires. According to their curriculum basic knowledge about IEC 61131-3 and UML was exalted.

*Design of the Study*

The experiment compared SysML-AT with a state of the art PLC programming language, i.e. Continuous Function Chart (CFC). CFC is a graphical language with graphic elements which are similar to the Parametric Diagram (PD). Even though CFC is not standardized by IEC 61131-3, it is provided by most IEC 61131-3 environments and widely used and accepted in industry for implementing closed loop control [86]. CFC provides free arrangement of Function Blocks (FB) and back coupling of FBs can be graphically realized [51]. There are only minimal differences, e.g. the free arrangement of FBs, to the Function Block Diagram (FBD) which is part of IEC 61131-3. According to Lüder et al. [87], other IEC 61131-3 languages like Ladder Diagram (LD) are less suitable for the design of (distributed) systems.

*Procedure*

The participants were divided into two groups according to the two notations (CFC and SysML-AT) and had to attend three different trainings. First, both groups attended an identical training that taught basic knowledge regarding automation technology. Subsequently, the groups were separated and two specialized trainings were conducted which taught the ideal-typical development approach using CFC or SysML-AT, respectively. The third training imparted practical skills in developing CFC and SysML-AT models. The overall training took half a day.

*Performance*

To compare SysML-AT and CFC regarding programming/ modeling performance the developed models/programs were stored electronically and analyzed manually by comparing the participants' models with a full master model. The performance was determined by the number of correctly modeled or programmed elements. To verify whether all requirements (R1–R5) are fulfilled by the presented approach, hypotheses (H1–H5) are used. Each hypotheses is verifies at least one requirement.

As given in Fig 15 and Fig. 16, the automation task included 4 PLCs, 7 actuators, 8 sensors, 10 non-functional requirements, and 5 different AFs which resulted in about 10--20 SAs. The figures show an extract of modeled solutions of two participants (CFC and SysML-AT). As the CFC solution shows, it is difficult to have an overview over the whole system (cf. Fig. 15). The SysML-AT addresses this problem (cf. Fig. 16) by combining the relevant information while providing a modeling tool with context-sensitive information highlighting and hiding.

*Hypothesis 1: Using SysML-AT leads to a better automation design (focusing entire model) (R1-R4)*

At the beginning of the experimental phase, requirement documents for a MS were handed out to the participants.

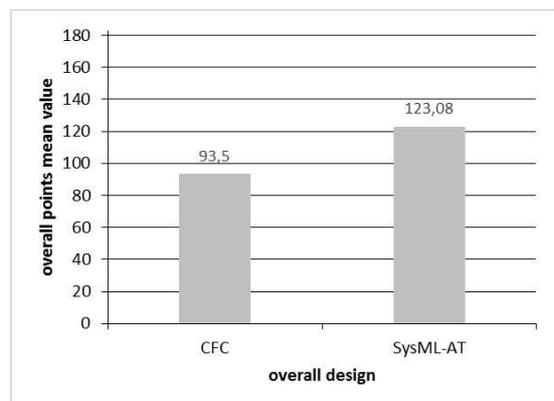

**Fig. 17: Participants' results CFC and SysML-AT (points ranging from 0 to 182pt)**

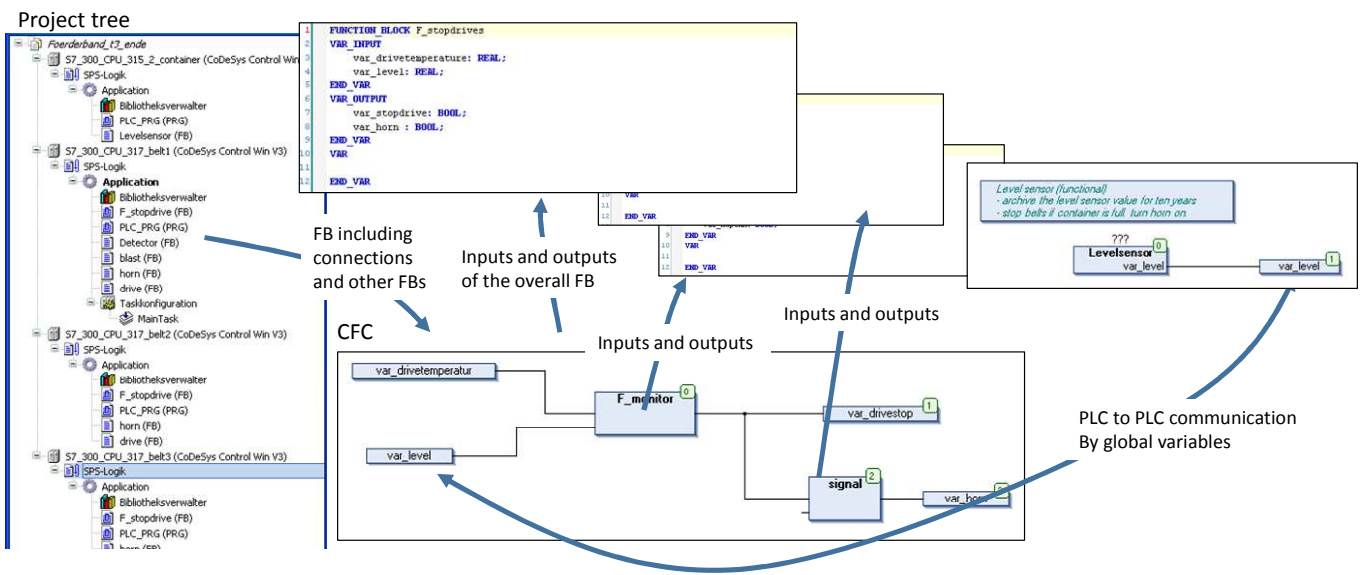

**Fig. 15: Extract of the best participant's model using CFC**

The models of the participants using SysML-AT showed a higher number of correctly modeled elements. Fig. 17 compares the mean value of points gained by the participants' solutions using either CFC or SysML-AT.

We assume a level of significance of p < 0.05, which means if probability that the hypothesis is wrong is smaller than 5% we assume the result as significant. Up to 10% we assume the result as trend. The experiments show a significant (probability of error/level of significance: p < 0.001) difference (see Fig. 17) to the group that used CFC. Hence, Hypothesis 1 is true.

*Hypothesis 2: Using SysML-AT leads to better software architecture (R2)*

As one task inside the experiment, the participants had to model the software and its behavior using SysML-AT or CFC. The identified functional requirements had to be decomposed to AFs and SAs. All necessary functions to operate the MS according to the given functional requirements had to be defined. The use of SysML-AT leads to more complete software models and, hence, better performance of the design regarding software architecture (mean values: CFC 17.30; SysML-AT 21.92; p = .054). Thus, Hypothesis 2 is considered valid.

*Hypothesis 3: Using SysML-AT leads to better hardware design and its properties (R3)*

In this task, the participants gathered all necessary hardware elements including their particular interfaces and properties (cf. Fig. 6). The group using CFC had to create nodes by selecting and naming PLCs from the library. To create a sensor or actuator they had to create a representing global variable using the correct data type.

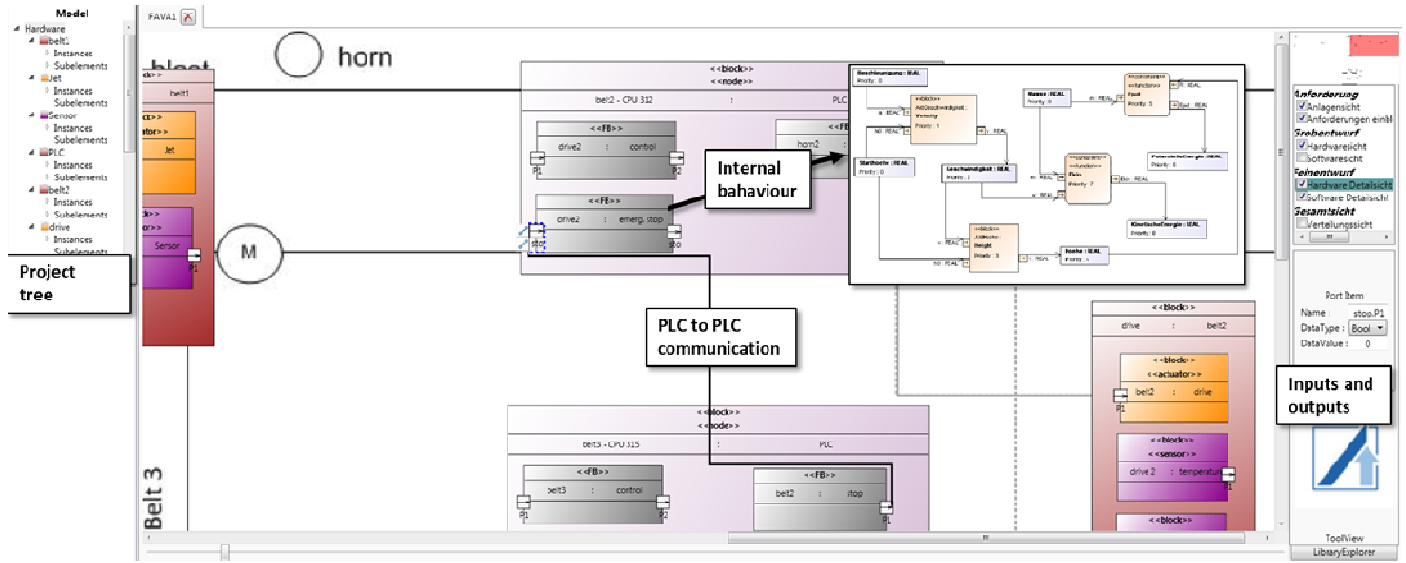

**Fig. 16: Extract of a participant's model using SysML-AT**

The SysML-AT group was able to use the provided diagrams and notations for sensors, actuators and nodes and describe the hardware interfaces by adding input and output ports. The explicit modeling supported by SysML-AT leads to better solutions (mean values: CFC 8.90; SysML-AT 15.00; p < 0,001) than classical table-based variable mapping. Thus, Hypothesis 3 can be considered true and R3 as fulfilled.

*Hypothesis 4: An integrated modeling of hardware and software (SysML-AT) leads to improved results in connection modeling (R4)*

The last task of the experimental design required the integration of the developed hardware and software models. Therefore, the defined SAs of the SysML-AT model or the FBs of the CFC software, respectively, had to be connected among each other and to the sensors and actuators. Since the interactions and connections between hardware-hardware, hardware-software and software-software are captured more completely and more correctly (mean values: CFC 10.40; SysML-AT 26.75; p < 0.001) if SysML-AT is used, Hypothesis 4 can be considered to be true.

*Hypothesis 5: SysML-AT provides better software maintenance (R5)*

The usability of the PD was evaluated by investigating the comprehensibility of model contents. The experiment compared notations with the same information content, i.e. PD, CFC and ST and required just model interpretation. All participants (N=6) were mechanical engineering students from different semesters. For the experiment, the permuted processing of three equally complex tasks was required. Each task demanded the participants to interpret a given model created in ST, CFC or PD.

The results indicated a positive tendency for the suitability of the PD, because the PD tasks were solved most completely (mPD = 68.25%) compared to the other two languages (mST = 64.28%, mCFC = 63.49%). It has to be recognized, that although the PD proved only to be marginally better in comparison to CFC as PLC programming languages the experiments indicated an equal or better suitability of this notation. Therefore, the PD can be assumed appropriate for modeling and maintaining software.

**6.3 Assessing the industrial applicability of SysML-AT**

Although the applicability and usability for laboratory manufacturing systems and mechanical engineering students could be proven, an additional survey was conducted among three industrial experts from different companies to assess the industrial applicability. Two of the interviewed experts were working for one of two major German vendors for automation technology (IEC 61131-3 programming environments, PLCs, sensors, drives, etc.) and one expert was working for a market leading machine and plant automation company.

As mentioned in [88] and recognized in [36] "*Surveys require considerable a priori knowledge of the phenomenon under study* [...] *the answerers would need to foresee the difficulties and benefits of a technology that they are not familiar with*" [36]. To alleviate this difficulty, industrial experts were chosen that on the one hand hold a PhD in engineering and have significant previous knowledge regarding MDE in automation but on the other hand have at least three years of industrial work experience. The interviews were conducted separately for each expert and comprised the presentation of the proposed MDE approach using a small consistent application example followed by the discussion of several interview questions. For answering the subsequently discussed interview questions, the experts had to explain their opinion on the suitability of the proposed method and the novel SysML-based notation.

All interviewed experts in general considered the presented MDE approach suitable and recognized the consistency of the approach (cf. R1 – R5) as the main benefit. Both experts from the automation technology vendors rated the proposed notation and method to map and connect software applications (SA) to physical hardware (nodes) (cf. R4) well suited for automation software development. This clearly underpins the results gained from the usability experiments presented in the previous subsection (cf. subsection 6.2, Hypothesis 4). However, it was further remarked that legacy automation software projects are often not function orientated but structured according to the mechanical structure of a MS. Hence, the possibilities for using pre-defined mechatronic components during the development process (cf. section 2) are considered highly relevant and have to be further elaborated for the MDE approach.

In a nutshell, the results obtained from the case studies and usability experiments can be underpinned by the experts' assessments of the MDE approach. Usability experiments that investigate the suitability of SysML-AT for professional automation engineers will be conducted in future works.

**7. Conclusions**

This paper presented an approach for the model-driven development of automation software for manufacturing systems. The approach is based on the Systems Modeling Language (SysML) and adapts the provided notations and meta-model to form a specialized language profile, the SysML-AT. For the MDE concept, five main requirements have been identified and an according modeling approach has been developed and evaluated.

The approach covers the modeling of functional and non-functional requirements. From described requirements, the functions that need to be implemented can be derived and captured within the same model. Hardware elements like sensors, actuators and nodes and their interfaces and properties are considered within the modeling approach as well. This enables the integration and linking of hardware and software models. The approach further comprises the model-based implementation and maintenance of automation software that complies with IEC 61131-3.

The newly developed SysML-AT provides language elements (stereotypes of SysML language elements) as well as diagrams and notations to describe the aforementioned aspects of a manufacturing system (MS). The prototypical tool-support that was realized for the application of SysML-AT implemented corresponding diagram editors and the graphical representations of the developed notation. To implement and maintain the developed automation software, an editor for SysML Parametric Diagrams has been integrated inside a widely used development and run-time environment for PLC-based automation software (CODESYS V3).

The realized integrated code generator enables the generation and implementation of code on the corresponding PLC run-time environment. In contrast to the related work, for the first time the modeling support was directly integrated inside a PLC software development and run-time environment providing an online-view with debugging features for the executed SysML-AT model. Another novelty of the approach is the explicit consideration of functional and non-functional requirements inside the model for manufacturing automation software combined with the modeling of the software applications and the automation hardware and its characteristics.

Several case studies using different types of manufacturing systems were conducted and proved the applicability of the MDE approach comprising the software modeling and code generation support. To evaluate the usability of the modeling approach, experiments with undergraduate mechanical engineering students have been conducted. To prove the positive influence on engineering results, the trials compared SysML-AT with a common PLC language (CFC) for control. The results showed that the developed MDE approach enables the development of better manufacturing automation software architectures. Hence, although additional engineering effort is required for the approach, e.g. for the explicit requirements modeling, this effort has been proven to get rewarded with better manufacturing automation software designs. Although the conducted usability experiments indicated positive effects of the proposed MDE approach and previous works of the authors have proven that comparable learning efforts are necessary for UML and IEC 61131-3 languages [89], further investigations to directly compare learning effort and achieved enhancements will be conducted. These works will investigate the applicability of the concept using industrial case studies.

The presented approach currently does not cover concepts for the validation of the system specification or the generated code, as e.g. presented in [90, 63, 28]. Hence, future research will be conducted that investigates possibilities for an automated validation of the system specification regarding the requirements described inside a SysML-AT model as well as the validation of the generated IEC 61131-3 code by automatically generated and executed software test cases (cf. [91]). Furthermore, the presented approach does not yet consider the integration of predefined software code during code generation. In future, we will integrate current work of the authors on automatic retrieval of already implemented software blocks [92]. To generalize the approach, the MDE will be extended to a more general definition of distributed systems. This may also include microcontroller-based nodes, robot controllers and numerical controllers. To achieve this, problems which are solved in PLC-based environments, like time synchronization and shared memory access, have to be addressed.